\documentclass{aa}
\usepackage{graphicx}
\usepackage{txfonts}

\begin{document}

\def\ltsima{$\; \buildrel < \over \sim \;$}
\def\simlt{\lower.5ex\hbox{\ltsima}}
\def\gtsima{$\; \buildrel > \over \sim \;$}
\def\simgt{\lower.5ex\hbox{\gtsima}}

\title{ Low-resolution spectroscopy \\ of the Sunyaev-Zel\'dovich effect \\
and estimates of cluster parameters}

\author{P. de Bernardis\inst{1,2}, S. Colafrancesco\inst{3,4},
G. D' Alessandro\inst{1}, L. Lamagna\inst{1,2}, \\ P.
Marchegiani\inst{3}, S. Masi\inst{1,2}, A. Schillaci\inst{1,2}}

\institute{ Dipartimento di Fisica, Universit\`{a} di Roma ``La
Sapienza", Roma, Italy
    \and INFN Sezione di Roma 1, Roma, Italy
    \and INAF - Osservatorio Astronomico di Roma, Monte Porzio Catone, Italy
    \and School of Physics, University of the Witwatersrand, Johannesburg Wits 2050, South Africa.}

\offprints{paolo.debernardis@roma1.infn.it}
\date{Submitted: Sept. 9$^{th}$, 2011 ; Accepted: Nov. 8$^{th}$, 2011}

{ \abstract {The Sunyaev-Zel\'dovich (SZ) effect is a powerful
tool for studying clusters of galaxies and cosmology. Large
mm-wave telescopes are now routinely detecting and mapping the SZ
effect in a number of clusters, measure their comptonisation
parameter and use them as probes of the large-scale structure and
evolution of the universe.}{We show that estimates of the physical
parameters of clusters (optical depth, plasma temperature,
peculiar velocity, non-thermal components etc.) obtained from
ground-based multi-band SZ photometry can be significantly biased,
owing to the reduced frequency coverage, to the degeneracy between
the parameters and to the presence of a number of independent
components larger than the number of frequencies measured. We
demonstrate that low-resolution spectroscopic measurements of the
SZ effect that also cover frequencies $> 270$ GHz are effective in
removing the degeneracy.}{We used accurate simulations of
observations with lines-of-sight through clusters of galaxies with
different experimental configurations (4-band photometers, 6-band
photometer, multi-range differential spectrometer, full coverage
spectrometers) and different intracluster plasma stratifications.}
{We find that measurements carried out with ground-based few-band
photometers are biased towards high electron temperatures and low
optical depths, and require coverage of high frequency and/or
independent complementary observations to produce unbiased
information; a differential spectrometer that covers 4 bands with
a resolution of $\sim 6 \ GHz$ eliminates most if not all bias;
full-range differential spectrometers are the ultimate resource
that allows a full recovery of all parameters. }{}}

\authorrunning{de Bernardis \emph{et al.}}
\titlerunning{Cluster parameters with SZ }
\maketitle

\section{Introduction}\label{introduction}

The Sunyaev-Zel\'dovich (SZ) effect (Sunyaev and Zeldovich,
\cite{Suny72}) is the inverse Compton energisation of cosmic
microwave background (CMB) photons that cross the hot plasma in
clusters of galaxies. The same effect is expected in other
astrophysical environments, such as the jets and lobes of giant
radio-galaxies (see e.g. Colafrancesco et al. \cite{Cola08}).

The SZ effect is a powerful tool for studying the physics of
clusters and using them as cosmological probes (see \emph{e.g.}
Birkinshaw \cite{Birk99}; Carlstrom et al. \cite{Carl02}; Rephaeli
et al. \cite{Reph06}).

Large mm-wave telescopes (Carlstrom et al. \cite{Carl11}; Swetz et
al. \cite{Swet11}; Schwan et al. \cite{Schw10}), coupled to
imaging multi-band arrays of bolometers, are now operating in
excellent sites and produce a number of detections and maps of the
SZ effect, discover new clusters, and establish cluster and
cosmological parameters.

Meanwhile, the Planck space mission (Planck collaboration
\cite{Plan11a}) has recently produced a shallow whole-sky survey
in nine cm to submm bands, from which an early catalogue of
massive clusters detected via the SZ effect has been extracted
(Planck collaboration \cite{Plan11b}). The Early-SZ catalogue
consists of 169 known clusters, plus 20 new discoveries, including
exceptional members (Planck collaboration \cite{Plan11c}).

All these measurements take advantage of the extreme sensitivity
of bolometers, with their excellent performance in the frequency
range 90-600 GHz where the spectral signatures of the SZ effect
lie.

Because all measurements integrate signals along the line of sight
(LOS), we summarize below the dependence of the LOS signal on the
physical parameters of the cluster. Several components should be
considered:

\smallskip \noindent
a) A thermal component. In the non-relativistic approximation for
the cluster plasma, the change of the CMB brightness $I$ in a LOS
crossing the cluster is [see e.g. Birkinshaw \cite{Birk99}]
\begin{equation}
{\Delta I_t \over I_{CMB}} = y {x^4e^x \over (e^x-1)^2} [ x
\coth(x/2) - 4 ]  ,  \label{sz}
\end{equation}
where $I_{CMB}$ is the brightness of the CMB, $x = h\nu / kT$ and
\begin{equation}
y = \int_{LOS} {kT_e \over m_e c^2} n_e \sigma_{T} d\ell ,
\label{yy}
\end{equation}
which is proportional to the integral of the pressure along the
LOS. For rich clusters the optical depth is $\tau_t = \int_{LOS}
n_e \sigma_{T} d\ell \simlt 0.01$ and the average energy boost of
CMB photons is $\Delta E / E \sim kT_e / m_e c^2 \simlt 1 \%$, so
that a $\Delta T / T \simlt 10^{-4}$ is expected. This is
reasonably large with respect to the intrinsic CMB anisotropy, and
has a characteristic spectrum, with a decrement of CMB brightness
at $\nu \simlt  218 GHz$ and an increment of CMB brightness at
$\nu \simgt  218 GHz$. This peculiar spectrum is the key to
distinguish the thermal SZ effect from competing contributions
(see below). Relativistic corrections have also been computed for
improved precision (see e.g. Itoh et al. \cite{Itoh00}) and are
used to characterize the physical parameters of the cluster
atmospheres (see e.g. Colafrancesco et al. \cite{Cola11}).

\smallskip \noindent
b) A Doppler component, caused by the collective motion of the
cluster with velocity $\rm{v}$ in the CMB restframe. The spectrum
of this effect is the same as the spectrum of the intrinsic CMB
anisotropy, so that the two contributions cannot be separated. The
amplitude of the signal is
\begin{equation}
{\Delta T \over T} \sim - \tau_t { \rm{v}_{LOS} \over c} ,
\end{equation}
where $\rm{v}_{LOS}$ is the projection of $\rm{v}$ along the LOS;
so we get
\begin{equation}
{\Delta I_{\rm{v}} \over I_{CMB}} \sim -\tau_t { \rm{v}_{LOS}
\over c} {xe^x \over (e^x-1)}  .
\end{equation}

\smallskip \noindent
c) A non-thermal component $\Delta I_{nt}$, caused by a
non-thermal population of electrons, produced by e.g. the AGNs
present in the cluster, relativistic plasma in cluster cavities,
shock acceleration. If dark matter in galaxy clusters consisted of
neutralinos, their annihilation would produce high-energy charged
particles as well. For a review, see e.g., Colafrancesco
\cite{Cola10} and references therein. The inverse-Compton spectrum
of this component is very different from that of the thermal
component, because the energy of CMB photons is boosted to
frequencies much higher than the submm frequencies observable with
CMB instruments (see Colafrancesco \cite{Cola08} for details).
This component is normally sub-dominant with respect to the
thermal component: its optical depth $\tau_{nt}$ is at least 50
times less than $\tau_t$. The parameters characterizing the
spectrum, in addition to the optical depth, are the spectral index
of the power-law spectrum of the energy of the electrons $\alpha$
(typically around -2.7), and their minimum momentum $p_1$,
typically of the order of a few $MeV/c$.

\smallskip \noindent
Additional sources of signal along the same line of sight are

\smallskip \noindent
d) the intrinsic anisotropy of the CMB (see point b) above)
\begin{equation}
{\Delta I_{CMBi} \over I_{CMB}} = {xe^x \over (e^x-1)} { \Delta T
\over T}   .
\end{equation}
Since components b) and d) have exactly the same spectrum, we will
describe them in the following with the parameter $\Delta I_{CMB}
= \Delta I_{CMBi} + \Delta I_{\rm{v}}$ (or the equivalent $\Delta
T_{CMB}$), which is the sum of the intrinsic anisotropy of the CMB
and of the kinetic SZ effect along the line of sight.

\smallskip \noindent
e) The emission of dust $\Delta I_d$ in our Galaxy and in the
galaxies of the cluster. This is modelled as a thermal spectrum
with temperature $T_d \sim 20K$ and a spectral index of emissivity
$\sim -1.5$, or a superposition of several components with
different $T_d$. It is important as a contaminant at frequencies
where the thermal SZ is positive.

\smallskip \noindent
f) The free-free and synchrotron emission ($\Delta I_{ff}$,
$\Delta I_{sy}$) from the diffuse medium in our Galaxy and from
the galaxies in the cluster: this component can be important at
low frequencies, where the SZ is negative.

According to these considerations it follows that SZ measurements
promise to estimate several physical parameters of the cluster on
the line of sight, provided there are more observation bands than
parameters to be determined, or some of the contributions are
known to be negligible.

\begin{figure}
\centering
\includegraphics[width=9cm]{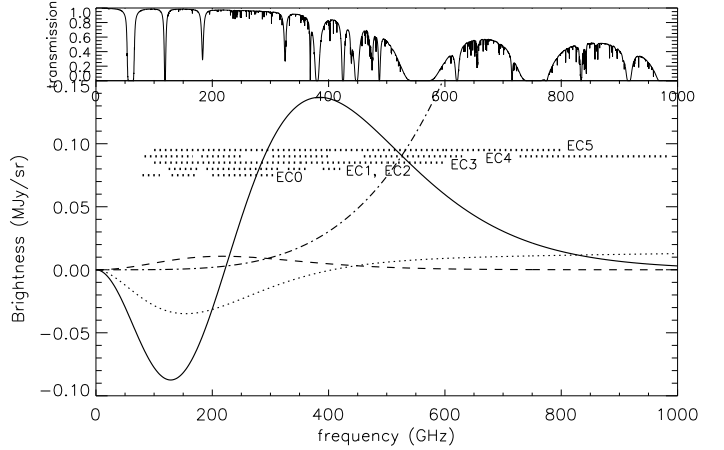}
\caption{Thermal SZ spectrum (continuous line in bottom panel),
compared to the atmospheric transmission of a dry, cold,
atmosphere (top panel, PWV=0.5mm), and to spectra of the
non-thermal SZ effect (dotted line), of CMB anisotropy and
kinematic SZ (dashed line), and of dust anisotropy (dot-dashed
line). The parameters of the different spectra are for the
benchmark case detailed in section \S \ref{sec2}: $\tau_t = 5
\times 10^{-3}$, $T_e = 8.5 keV$, $\Delta T _{CMB}=22 \mu K$,
$\tau_{nt} = 1 \times 10^{-4}$, $\alpha = -2.7$, $p_1 = 1.4
MeV/c$, $\Delta I_d (150\ GHz)=600 \ Jy$. The frequency coverage
of the different experiments considered in the paper is also shown
as dotted horizontal lines, labelled with the experimental
configuration number. \label{fig1} }
\end{figure}

Multi-frequency measurements are therefore mandatory to separate
the contributions of the different physical components, taking
advantage of the characteristic spectrum of the SZ effect, which
significantly departs from the spectra of the foreground and
background components (see fig. \ref{fig1}). The wider and more
detailed the frequency coverage of the observations, the more
effective the separation of the different components.

The recent results of Planck (Planck collaboration
\cite{Plan11b}), for example, have been obtained by exploiting the
excellent frequency coverage of the mission and sophisticated
component separation techniques (see {\it e.g.} Leach et al.
\cite{Leac08}).

In this paper we study how effective the various experimental
configurations are in separating all the different physical
components and in providing unbiased estimates of the cluster
parameters (like $y$, $\rm{v}$, $T_e$, $\tau_t$, $\tau_{nt}$,
$p_1$ ...) and of the other parameters that describe the signals
along the same line of sight ($T_d$, $\tau_d$, $I_s$, $I_{ff}$,
$\Delta I_{CMBi}$, ...).

Evidently, ground-based few-band photometers cannot provide enough
information to separate all physical components. Observations are
hampered at high frequencies ($\simgt 200 GHz$) by atmospheric
noise (see fig.\ref{fig1}): this significantly limits the coverage
of the positive part of the thermal SZ spectrum, and makes the
removal of parameter degeneracies much more difficult. These
instruments need external information (optical, X-ray, far-IR,
etc.) to produce mainly measurements of $\tau_t$. With the
addition of external data, these experiments provide invaluable
information in the current exploration phase: a large database of
clusters is being built, and new cluster candidates have been
discovered (see e.g. Hincks et al. \cite{Hinc10}; Marriage et al.
\cite{Marr10}; Brodwin et al. \cite{Brod10}; Hand et al.
\cite{Hand11}; Sehgal et al. \cite{Sehg11}; Foley et al.
\cite{Fole11}; Story et al. \cite{Stor11};  Williamson et al.
\cite{Will11}).

In principle, future space-based spectrometers can cover the full
range of interesting frequencies and offer much more information:
with these machines it should be possible to measure the
parameters of a cluster, and use external information, when
available, as a cross-check. Also, other important scientific
targets of these instruments are the measurement of the $C^+$ and
CO lines in the redshift desert and beyond for a large number of
galaxies (see e.g. de Bernardis et al. \cite{debe10}; Gong et al.
\cite{Gong11}).

The main purpose of this paper is to analyse quantitatively and
compare the performance of these different configurations for SZ
measurements.

As samples of experimental configurations we considered a
ground-based 4-band photometer; a 4-band photometer and a 4-band
differential spectrometer (both suitable for balloon-borne
observations, and similar to the configuration of the OLIMPO
experiment - see Masi et al. \cite{Masi08}; Conversi et al.
\cite{Conv10} ), Planck HFI (Planck HFI core team \cite{Plan11d}),
and a full-coverage differential spectrometer suitable for a
future, dedicated space mission, like SAGACE (de Bernardis et al.
\cite{debe10}) or Millimetron
(http://www.sron.rug.nl/millimetron).

In section \S \ref{sec2} we describe and compare the considered
configurations; in section \S \ref{sec3} we describe the
simulations and the analysis method; in section \S \ref{sec4} we
discuss the results.

\section{Experimental configurations}\label{sec2}

All instruments considered here work at the diffraction limit,
with $A \Omega = \lambda^2$, to achieve the best possible angular
resolution, which is needed to resolve the target cluster.

We limited our analysis to experiments using arrays of bolometers
because of their superior mapping speed. Having maps of a sky
patch surrounding the cluster is important to identify foreground
structures and the internal structure of the cluster, if the
resolution is sufficient. This choice limits the frequency
coverage to $\nu \simgt 70 GHz$, where bolometer arrays have a
significant sensitivity advantage with respect to coherent
detectors.

Few-band photometers have been traditionally matched to the
mm/submm atmospheric windows, i.e. the frequency bands
W1=[75-115]GHz, W2=[125-175]GHz, W3=[190-315]GHz, W4=[330-365]GHz,
W5=[390-420]GHz. Operation in the higher frequency windows is
significantly hampered by atmospheric noise and poor transmission,
which strongly depends on the telescope site. In general
ground-based photometers use narrower bands within the ranges
above, while balloon-borne multi-band photometers can have wider
bands and can use the higher frequency bands efficiently.
Satellite instruments do not have these limitations, and the
operation bands are selected based on the necessity to calibrate
the instrument and to study the Galactic and extragalactic
contaminating foregrounds (like the CO lines, and the continuum
from interstellar dust, free-free and synchrotron emission).

For our spectroscopy we considered a Fourier Transform
Spectrometer (FTS) because, at variance with dispersion
spectrometers (see e.g. Bradford et al. \cite{Brad02} ), this
instrument is intrinsically imaging, a crucial requirement for
studying appropriately the SZ effect in cosmic structures.
Moreover, at variance with Fabry-Perot spectrometers (see e.g.
Benford et al. \cite{Benf02} ) the FTS can be used in a
differential configuration where a sky field is compared to an
internal reference blackbody (as in the COBE-FIRAS instrument,
Mather et al. \cite{Math93}, or in the Herschel-SPIRE, Griffin et
al. \cite{Grif07}), and also in a differential configuration where
two sky-fields are compared, thus rejecting most of the common
mode signal from the instrument, the atmosphere, the foregrounds,
and the CMB (see e.g. de Bernardis et al. \cite{debe10}).

Experimental Configuration 0 (EC0) is a ground-based 4-band
photometer that measures the bands W1, W2, and the two halves of
W3 ([200-240]GHz and [240-310]GHz) in an excellent site like the
Atacama desert, South Pole, or Dome-C. We assumed 2.7\%, 2.5\%,
3.5\%, 5.0\% average emissivity for the atmosphere, respectively,
in the 4 bands above (corresponding to a precipitable water vapor
$\sim 0.5$ mm), and an equivalent temperature of the atmosphere of
240K. Only photon noise was considered, assuming that the
telescope is used in a differential configuration, where most of
the turbulence is subtracted out as a common mode signal.

Experimental Configuration 1 (EC1) is a balloon-borne 4-band
photometer that measures bands W2, W3, W4, and W5. Here the
background is limited by the temperature ($\sim$ 230 K) and
emissivity of the telescope mirrors and of the cryostat window.
Again a differential configuration was considered.

Experimental Configuration 2 (EC2) is a balloon-borne 4-band
differential spectrometer that measures the low-resolution
($\Delta \nu = 6$ GHz) spectra in four ranges coincident with W2,
W3, W4, and W5. Here the background is limited by the temperature
and emissivity of the telescope, of the interferometer mirrors,
and of the cryostat window, all at room temperature (230K). Note
that in this (and in the following) photon-noise limited
configurations, that use Fourier transform spectrometers, the
error on each spectral bin scales as the inverse of the spectral
resolution $\Delta \nu$ (see \S \ref{sec3}).

Experimental Configuration 3 (EC3) is an Earth-orbit (EO)
satellite with a telescope radiatively cooled at 80K, which
measures spectra in four consecutive ranges b0 = [100, 200]GHz, b1
= [201, 350]GHz, b2 = [351, 500]GHz, b3 = [501, 600]GHz with 6 GHz
resolution. Here the background is limited by the temperature and
emissivity of the cold telescope and of the cryostat window (total
emissivity $\epsilon \sim 0.01$), while the interferometer mirrors
are kept at $<4$K and do not contribute to the photon noise. The
spectral coverage has been divided in the four ranges observed
simultaneously by independent detector arrays in order to limit
the radiative background on each array.

Experimental Configuration 4 (EC4) is a 6-band photometer similar
to Planck-HFI, which operates in the Lagrangian point L2 of the
Sun-Earth system, with the telescope radiatively cooled at 45K,
measuring simultaneously in the bands b0 = [83.5, 116.5]GHz,  b1 =
[119.4, 166.6]GHz, b2 = [181.2, 252.8]GHz, b3 = [303.6, 402.4]GHz,
b4 = [460.5, 629.5]GHz, b5 = [728.5, 985.6]GHz. Here the
background is limited by the temperature and emissivity of the
cold telescope (total emissivity $\epsilon \sim 0.005$).

Experimental Configuration 5 (EC5) is a 4-range differential
interferometer that operates in deep space (L2), with an actively
cooled telescope ($T \simlt 6 K$), in the bands b0 = [100,
200]GHz, b1 = [201, 400]GHz, b2 = [401, 600]GHz, b3 = [601,
800]GHz with 6 GHz resolution. Here the background is limited by
natural radiation sources (Galactic, extragalactic, CMB): this
allows a wider coverage (including high frequencies) without the
risk of a high background on the detectors.

For a summary see fig.\ref{fig1}, where the coverage of the
different configurations is compared to the SZ and foreground
spectra.

\section{Simulation of line-of-sight observations}\label{sec3}

In order to evaluate the performance of different experimental
configurations, we have considered the following benchmark
situation: the observation of a line of sight that crosses a rich
cluster of galaxies, with $\tau_t = 0.005$, $T_e = 5 keV$, $\Delta
T_{CMB} = 22 \mu K$ (corresponding to $ \rm{v}=480 km/s $, if the
intrinsic anisotropy of the CMB along the line of sight is
negligible), $\tau_{nt} = 0.0001$, $\alpha=-2.7$, $p_1 = 2.75
\times 511 keV/c$).

We assumed that the angular resolution of all channels of all
experimental configurations is sufficient to resolve the source.
Otherwise we would have needed to take into account dilution and
shape factors: in this LOS approach we avoided these
complications. We believe that this approach is adequate for the
purpose of this paper, which is to compare the performance of
different experimental configurations.

The integration time on the same line of sight is assumed to be 3
hours for all cases but EC4. For EC4 (a space-borne whole-sky
survey {\it \'a-la-Planck}) the integration time on the considered
sky pixel is 30 s.

The signal power on the detectors was computed as
\begin{equation}
S(\nu) = A\Omega E(\nu)(1-\epsilon_m(\nu)) \left[ \Delta I_t +
\Delta I_{\rm{v}} + \Delta I_{CMB} + \Delta I_{nt} + \Delta I_d
\right] , \label{theory}
\end{equation}
where $A$ is the collecting area, $\Omega$ is the solid angle
sampled by each detector, $E(\nu)$ is the efficiency of the
detection system, $\epsilon_m(\nu)$ is the total emissivity of the
optical system at room temperature and of the atmosphere in the
measurement band.

We have neglected  $\Delta I_{ff}$ and $\Delta I_{sy}$ because
they are negligible with respect to $\Delta I_d$ in the frequency
range and for the observations at high galactic latitudes of
interest here. We modelled typical Galactic cirrus anisotropy at
the angular scale of a cluster (a few arcmin) as $\Delta I_d (\nu)
= A (\nu / \nu_o)^4 $ with $A = 600 \ Jy/sr$ and $\nu_o = 150 \
GHz$: a value typical of very clean high-latitude regions (see
e.g. Masi et al. \cite{Masi06}).

We assumed that the detector array is optimized to be limited by
the photon noise of the radiative background that is produced by
the instrument and the atmosphere (if present). Cryogenic
bolometers reach this performance level even in the extremely low
background achievable in space, if properly designed (see e.g.
Holmes et al. \cite{Holm08}).

The background power on the detectors was computed as
\begin{equation}
B(\nu) = A \Omega E(\nu) \left[ \epsilon_m(\nu) {2h\nu^3 \over
c^2}{1 \over e^{x_m}-1} + \left[1-\epsilon_m(\nu)\right]{2h\nu^3
\over c^2}{1 \over e^x-1} \right] , \label{background}
\end{equation}
where  $x_m = h\nu / kT_m$, $T_m$ is the temperature of the
optical system and of the atmosphere, $x = h\nu / kT$, $T$ is the
temperature of the cosmic microwave background.

The fluctuations of the background were computed as

\begin{equation}
NEP_{ph}^2 = NEP_m^2 + NEP_{CMB}^2  \label{noise}
\end{equation}

\begin{equation}
NEP_m^2 = A \Omega {4 k^5 \over c^2 h^3}  T_m^5 E(\nu)
\epsilon_m(\nu) {x_m^4 \left[e^{x_m}-1+ E(\nu)
\epsilon_m(\nu)\right] \over (e^{x_m}-1)^2 }
\end{equation}

\begin{eqnarray}
 NEP_{CMB}^2  =  &  & \nonumber  \\
 = A \Omega {4 k^5 \over c^2 h^3}  T^5 E(\nu)
(1-\epsilon_m(\nu))  {x^4 \left[e^{x}-1+ E(\nu)
(1-\epsilon_m(\nu))\right] \over (e^{x}-1)^2 } & &
\end{eqnarray}

For the photometric measurements we integrated equations
\ref{theory} and \ref{noise} over the detection bandwidth $BW$ to
obtain the error on the signal for each band:
\begin{equation}
\sigma_{phot}^2 = {\int_{BW} NEP_{ph}^2 d\nu \over {2T}} ,
\label{noise_phot}
\end{equation}
where $T$ is the total integration time.

In the case of spectroscopic measurements, the error on the
measurement of each spectral bin was computed as follows. In the
FTS the input power is splitted in the two arms of the
interferometer, and a variable delay is introduced in one of the
two beams, before recombining them on the detector. The variable
delay is introduced by a moving mirror, which can be offset by $x$
with respect to the corresponding steady mirror in the other arm
of the interferometer. In this way a $2cx$ delay is introduced.
The power measured in position $x$ of the moving mirror is $P(x)$
(the interferogram), and the spectrum is estimated as the Fourier
transform of the interferogram :
\begin{eqnarray}
S(\sigma) = \int_{-x_{max}}^{x_{max}} (P(x)-\langle P \rangle)
\cos (4\pi \sigma x) dx = & & \nonumber \\ = 2 \int_0^{x_{max}}
(P(x)-\langle P \rangle) \cos (4\pi \sigma x) dx , & &
\end{eqnarray}
where $\sigma$ is the wavenumber (in cm$^{-1}$). The spectral
resolution of the measurement is
\begin{equation}
\Delta \sigma = 1.22/(2 x_{max})
\end{equation}
(see e.g. Chantry \cite{Chan71}). In a real instrument the
interferogram is sampled at the positions $x_i=i\Delta x$,
$(i=1..N)$, with $N=x_{max}/\Delta x$. The integral is then
estimated as a discrete sum
\begin{eqnarray}
S(\sigma_j) = & 2 \sum_{i=1}^N (P(x_i)-\langle P \rangle) \cos
(4\pi \sigma_j x_i) \Delta x =  & \nonumber \\ & = 1.22
\sum_{i=1}^N (P(x_i)-\langle P \rangle) \cos (4\pi \sigma_j x_i) /
( N \Delta \sigma) . &
\end{eqnarray}
Since each sample of the interferogram has an integration time
$T/N$, and photons from the whole observed frequency bandwidth
contribute to the noise, the error in the measurement is therefore
\begin{equation}
\sigma_{P} = {\sqrt{ \int_{BW} NEP_{ph}^2 d\nu } \over
\sqrt{2T/N}} .
\end{equation}
The error in the measurement of each spectral bin is thus
\begin{eqnarray}
\sigma^2_S = & 1.22^2 \sum_{i=1}^N  \sigma^2_{P} \cos^2 (4\pi
\sigma x_i) / ( N \Delta \sigma)^2 =  & \nonumber \\ & = 1.22^2
{\sigma^2_{P} \over N \Delta \sigma^2 }  \sum_{i=1}^N {\cos^2
(4\pi \sigma x_i) \over N} \sim 0.74 {\sigma^2_{P} \over N \Delta
\sigma^2 } , &
\end{eqnarray}
or
\begin{equation}
\sigma_S = 0.61 {\sqrt{\int_{BW} NEP^2_{ph} d\nu} \over \Delta
\sigma \sqrt{T}} = 0.61 {\sqrt{\int_{BW} NEP^2_{ph} d\nu } \over c
\Delta \nu \sqrt{T}} . \label{noise_spec}
\end{equation}

For each experimental configuration (EC*) we simulated 3000
measurements, using equation \ref{theory}, and adding an error
term extracted from a Gaussian distribution with zero average and
standard deviation derived from equations \ref{noise_phot} and
\ref{noise_spec}.

\begin{figure}
\centering
\includegraphics[width=9cm, height=6cm]{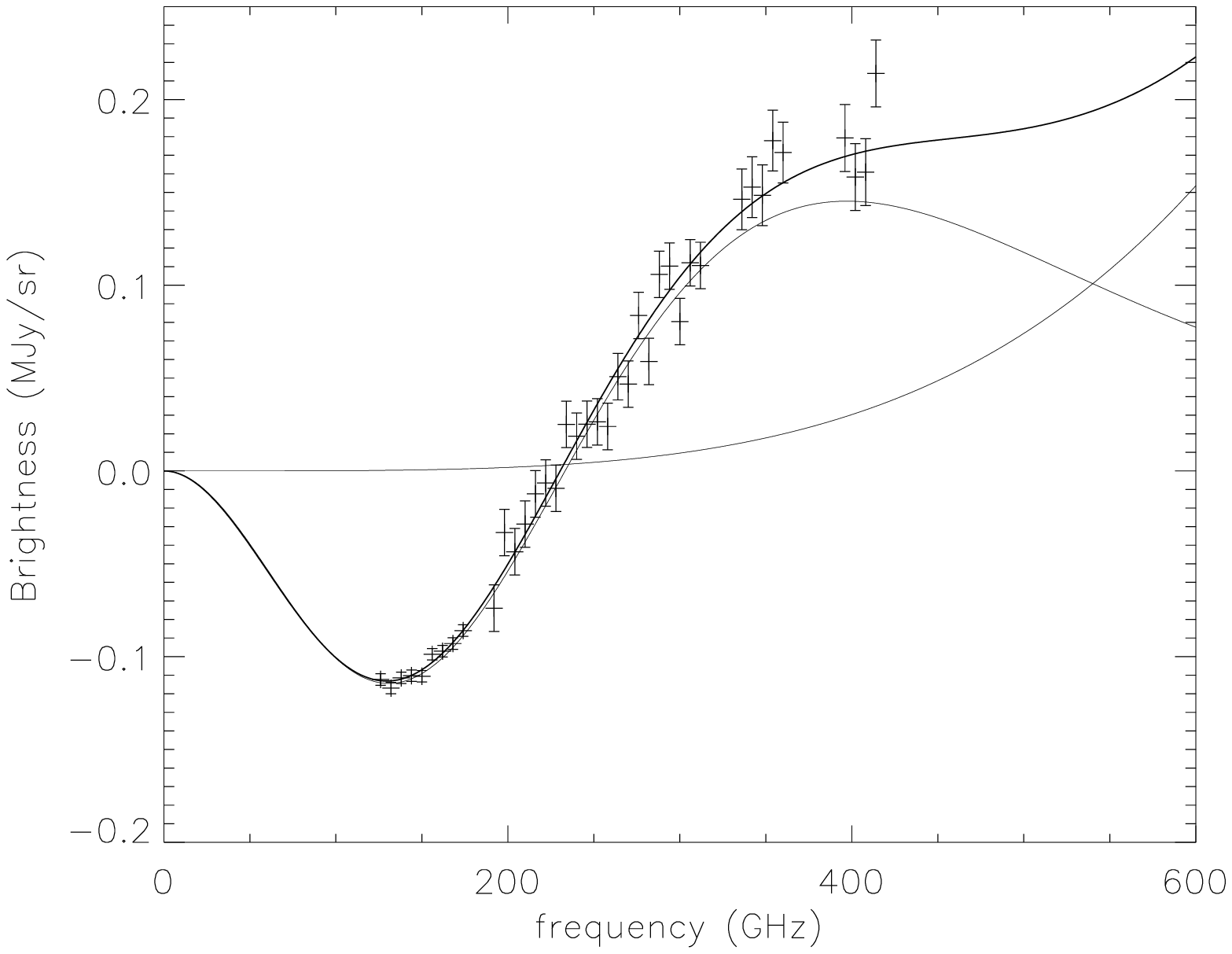}
\includegraphics[width=9cm, height=6cm]{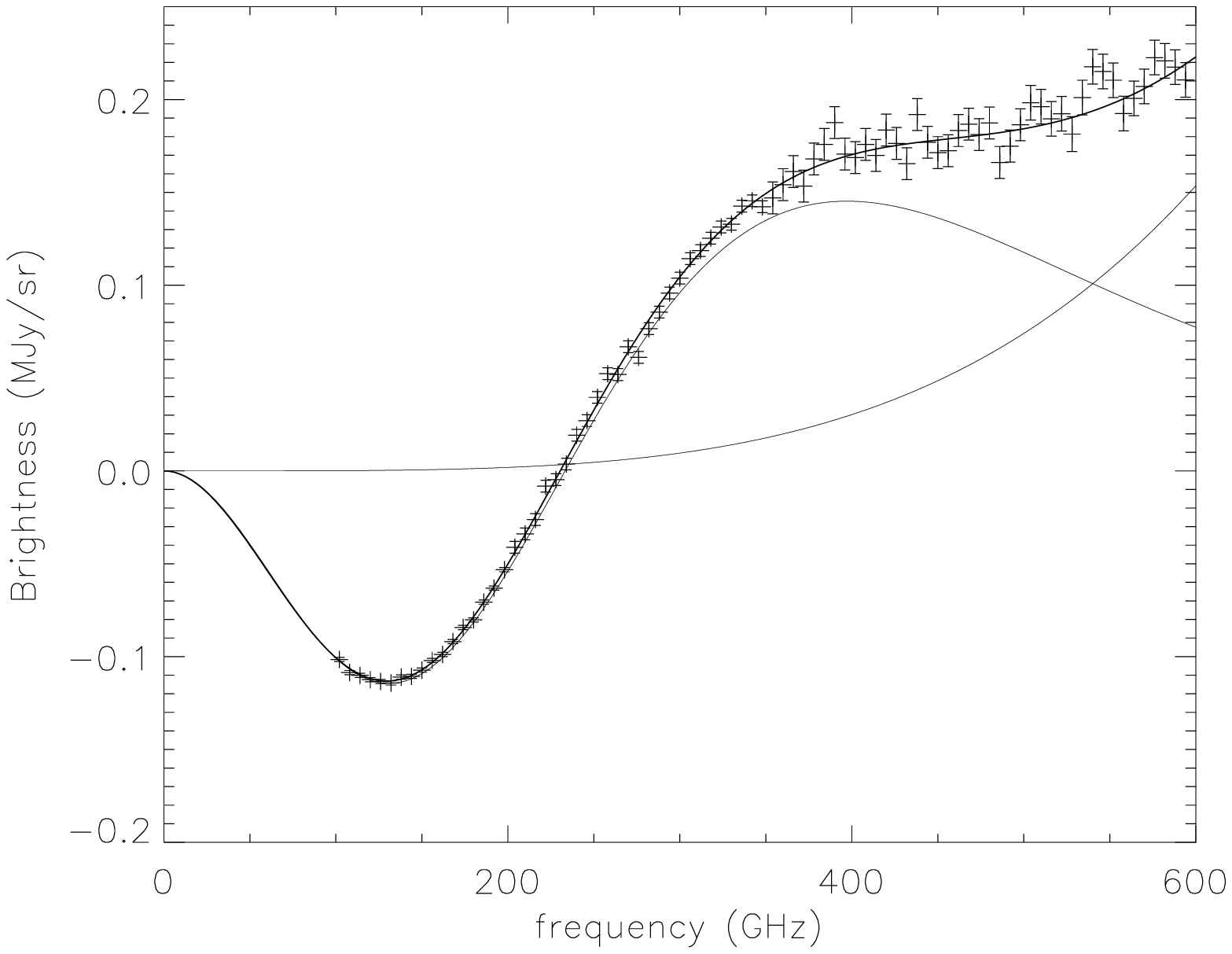}
\includegraphics[width=9cm, height=6cm]{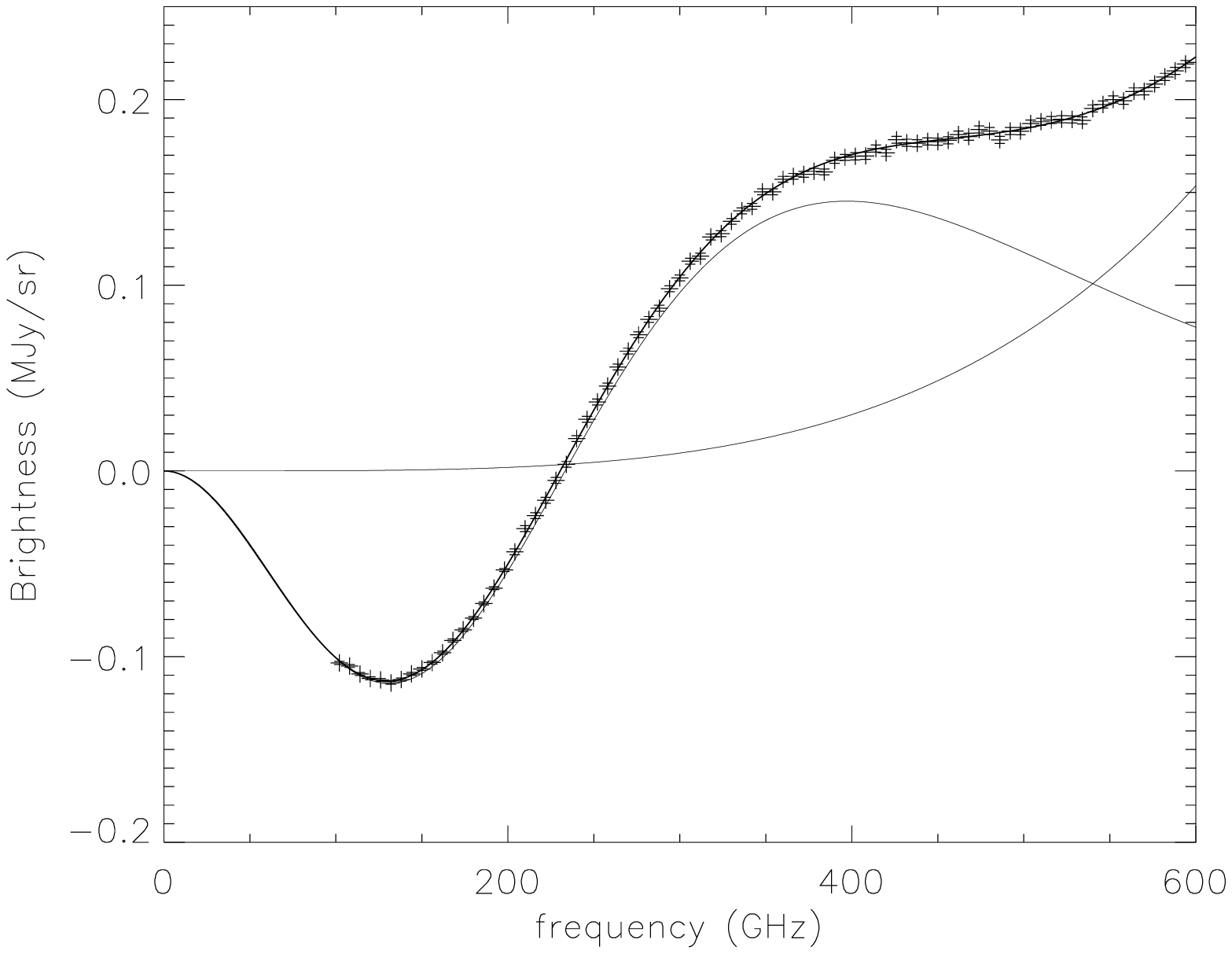}
\caption{From top to bottom: simulated data sets for the
spectroscopic configurations EC2 (differential FTS on a
stratospheric balloon, room temperature), EC3 (differential cold
FTS on an earth-orbit satellite, with room temperature telescope),
EC5 (differential cold FTS on a L2-orbit satellite, with cold
telescope). The best-fit line through the data points is from eq.
\ref{theory}. The other two lines are the thermal plus non-thermal
SZ, and the dust anisotropy. \label{fig2} }
\end{figure}

Typical simulated spectral measurements are reported in figure
\ref{fig2}.

\section{Results and discussion}\label{sec4}

We fitted each simulated measurement using equation \ref{theory}.
In table \ref{tab1} we report the averages of the best-fit
parameters with their standard deviation.

\begin{figure}
\centering
\includegraphics[width=9cm, angle=90]{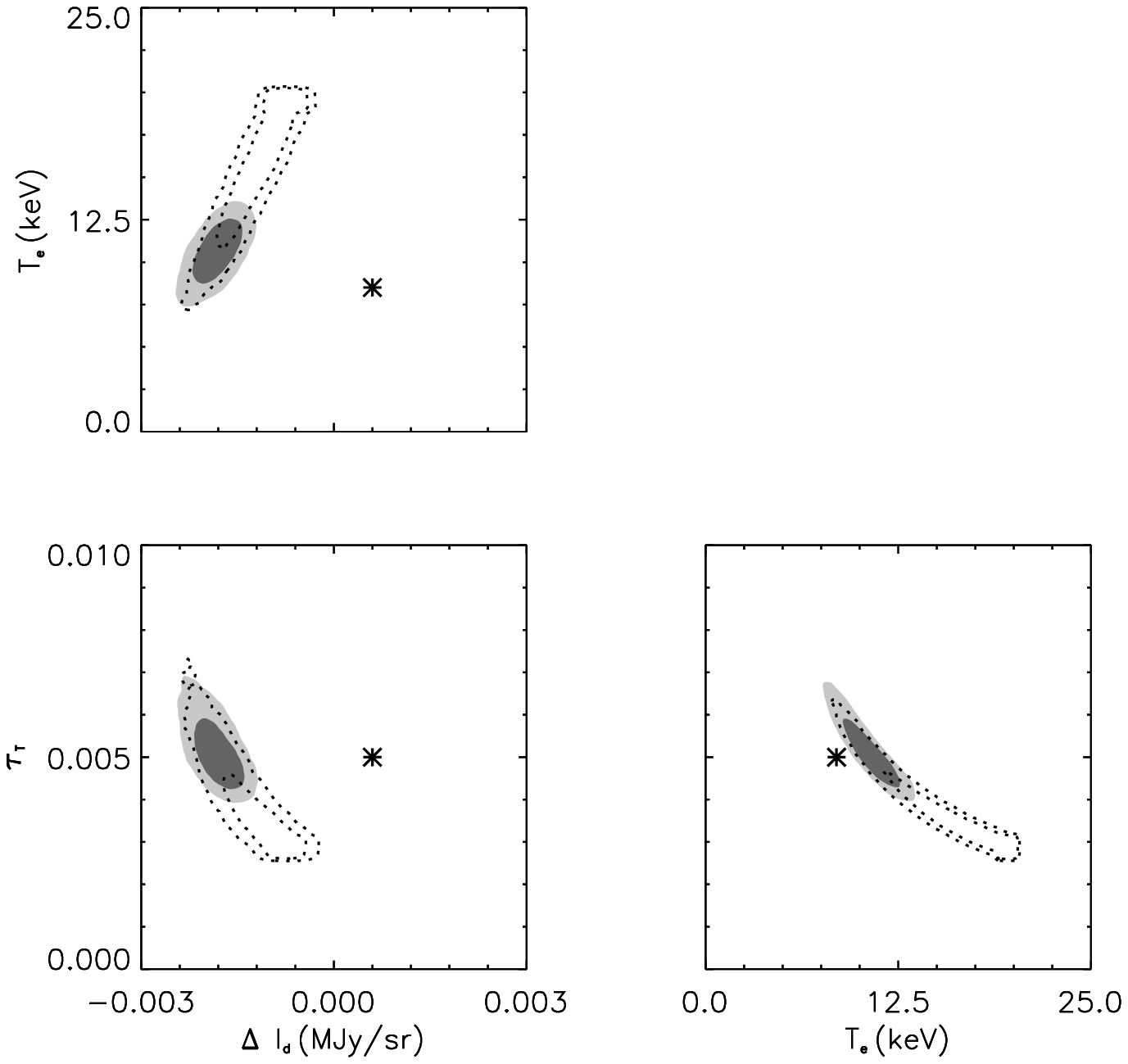}
\caption{ Joint likelihood contours (95.6$\%$ and 68.7$\%$) for
couples of best-fit parameters for the experimental configuration
EC0 (ground based observations with a 4-band photometer) where the
data are analysed by fitting the three parameters $\tau_t$, $T_e$,
$\Delta I_d$. The filled contours are for a Gaussian prior on
$T_e$ with $\sigma = 3 \ keV$; the dashed contours are for a
Gaussian prior with $\sigma = 8 \ keV$. The * symbols mark the
input values of parameters. In this case a significant bias for
all parameters is evident. The anti-correlation between $\tau_t$
and $T_e$ is also evident. \label{fig3a} } \centering
\includegraphics[width=11cm, angle=90]{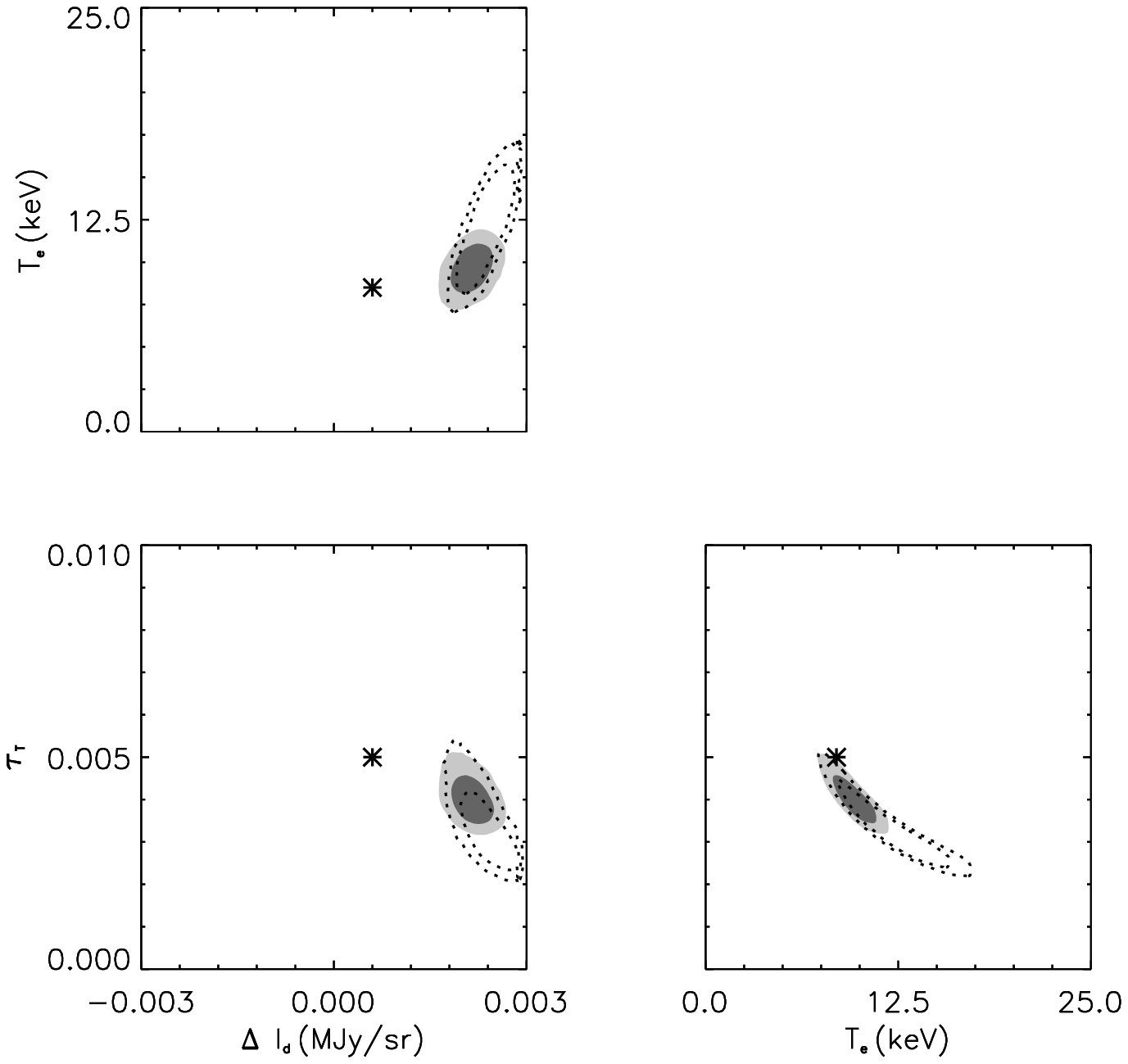}
\caption{ Same as fig. \ref{fig3a} but with $\tau_{nt}=0$. The
bias changes, but does not vanish.  \label{fig3b} }
\centering
\includegraphics[width=11cm, angle=90]{spettri_EC00_3_all_gs.eps}
\caption{ Same as fig. \ref{fig3a} but with $\tau_{nt}=0$. The
bias changes, but does not vanish.  \label{fig3b} }
\end{figure}

\begin{figure}
\centering
\includegraphics[width=9cm, angle=90]{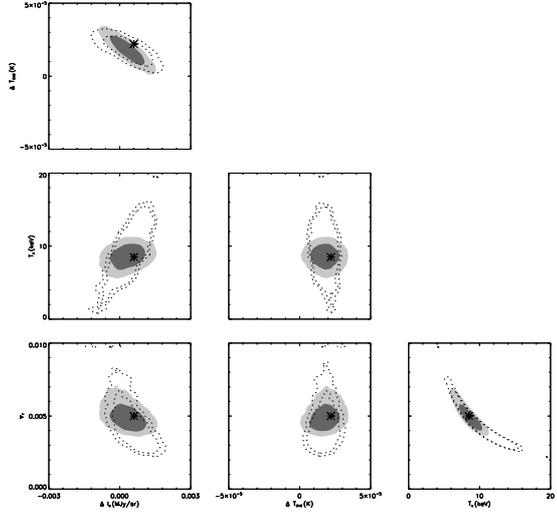}
\caption{ Same as fig. \ref{fig3a} when the data are analysed by
fitting four parameters ($\tau_t$, $T_e$, $\Delta I_d$ and $\Delta
T_{CMB}$). The bias is reduced, but the anti-correlation between
$\tau_t$ and $T_e$ is still evident and the confidence contours
are wider. \label{fig4} }
\end{figure}

\begin{figure}
\centering
\includegraphics[width=9cm, angle=90]{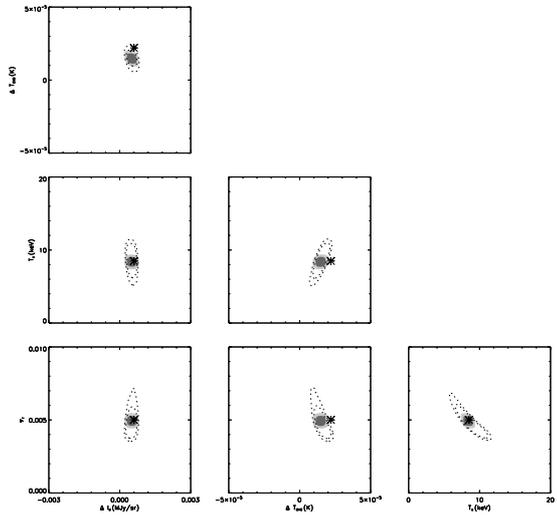}
\caption{ Joint likelihood contours (95.6$\%$ and 68.7$\%$) for
couples of best-fit parameters for the experimental configuration
EC1 (balloon-borne observations with a 4-band photometer,
extending the coverage to high frequencies not observable from the
ground) where the data are analysed by fitting four parameters
($\tau_t$, $T_e$, $\Delta I_d$, $\Delta T_{CMB}$). The filled
contours are for a Gaussian prior on $T_e$ with $\sigma = 3 \
keV$; the dashed contours are for a Gaussian prior with $\sigma =
8 \ keV$. The * symbols mark the input values of parameters.
\label{fig5} }
\end{figure}

While giving a general idea of the relative efficiency of the
different configurations, the results reported in table \ref{tab1}
can be misleading in the details, since the distributions of the
best-fit parameters are not Gaussian (nor symmetrical), and there
are significant correlations between the parameters. This is
evident from the joint likelihood contours plotted in figures
\ref{fig3a}, \ref{fig3b}, \ref{fig4}, \ref{fig5}, \ref{fig6},
\ref{fig7}, \ref{fig8}.

For EC0 (4-band ground-based photometer), where only four
independent data-sets are available for each measurement, we tried
to fit either three parameters ($\tau_t$, $T$, $\Delta I_d$) or
four parameters ($\tau_t$, $T_e$, $\Delta I_d$ and $\Delta
T_{CMB}$ ), adding a fictitious data point with zero brightness at
zero frequency. The results are dominated by the degeneracy
between $T_e$ and $\tau_t$, evident from equations \ref{sz} and
\ref{yy}: without relativistic corrections, the thermal SZ depends
on the product of electron density and electron temperature. For
this reason a decrease of, say, a factor 2 of $\tau_t$ is almost
perfectly compensated for by an increase of a factor 2 of $T_e$.
The only way to break the degeneracy is through the relativistic
corrections, which, however, are very small: their effect is
negligible with respect to the typical uncertainties of
ground-based measurements. For this reason we had to use a prior
on $T_e$, assuming that the information is obtained through
independent X-ray measurements of the specific brightness of the
cluster along the same line of sight. We tried a very weak prior,
with a Gaussian distribution centred on the true value and a
standard deviation of 8 keV: otherwise the best fit would converge
to non-physical values for $T_e$. The bias is only mitigated by
the introduction of this prior. In the 3-parameter case the best
fits converge on values of the parameters that are not very close
(in units of their standard deviation) to the input values, and
the typical $\chi^2$ is high, confirming that the effect of the
measurement error is negligible with respect to the effect of the
parameter degeneracies and to the necessity of neglecting the
non-thermal component in the fit.

In figure \ref{fig3a} we plot the joint likelihood contours for
couples of parameters in the 3-parameter fits. The comparison with
figure \ref{fig3b} shows that part but not all of the bias depends
on the presence of the non-thermal component: indeed the bias
changes with $\tau_{nt}=0$, but is still present.

If a 4-parameter fit is used, the bias is significantly reduced,
and does not depend anymore on the presence of the non-thermal
component. However, this comes at the cost of wider confidence
areas, as evident from figure \ref{fig4}.

The contamination of dust is significant and degenerates to some
extent also with $\Delta T_{CMB}$: in the 3-parameter fit, where
$\Delta T_{CMB}$ is not fitted, the best-fit for $\Delta I_d$ is
far from the input value, and this probably contributes to the
tension between the best-fit value of $T_e$ and its input value.
We believe that this is the result of a degeneracy of parameters
combined with the poor coverage of high frequencies in
ground-based experiments. In the 4-parameter fits the $\chi^2$
improves, but the best fits are still biased, and only with the 3
keV prior the estimates of all parameters are consistent with the
input values (see fig. \ref{fig4}).

For EC1 (4-bands balloon-borne photometer) the coverage of
frequencies higher than 300 GHz helps in removing the degeneracy
(see fig. \ref{fig5}). If we use the very weak prior on $T_e$
(Gaussian with standard deviation 8 keV), the best-fit cluster
parameters are already close to the input values, even if the
error (completely dominated by the degeneracy) is relatively
large. The situation improves, of course, if the standard
deviation on the prior is reduced to 3 keV. $\Delta T_{CMB}$ is
better constrained, but is biased low. While performing
significantly better than EC0, this configuration is limited by
radiation noise in the high-frequency bands, which can be removed
only by cooling the telescope or/and adding spectroscopic
capabilities (see below).

\begin{figure}
\centering
\includegraphics[width=9cm, angle=0]{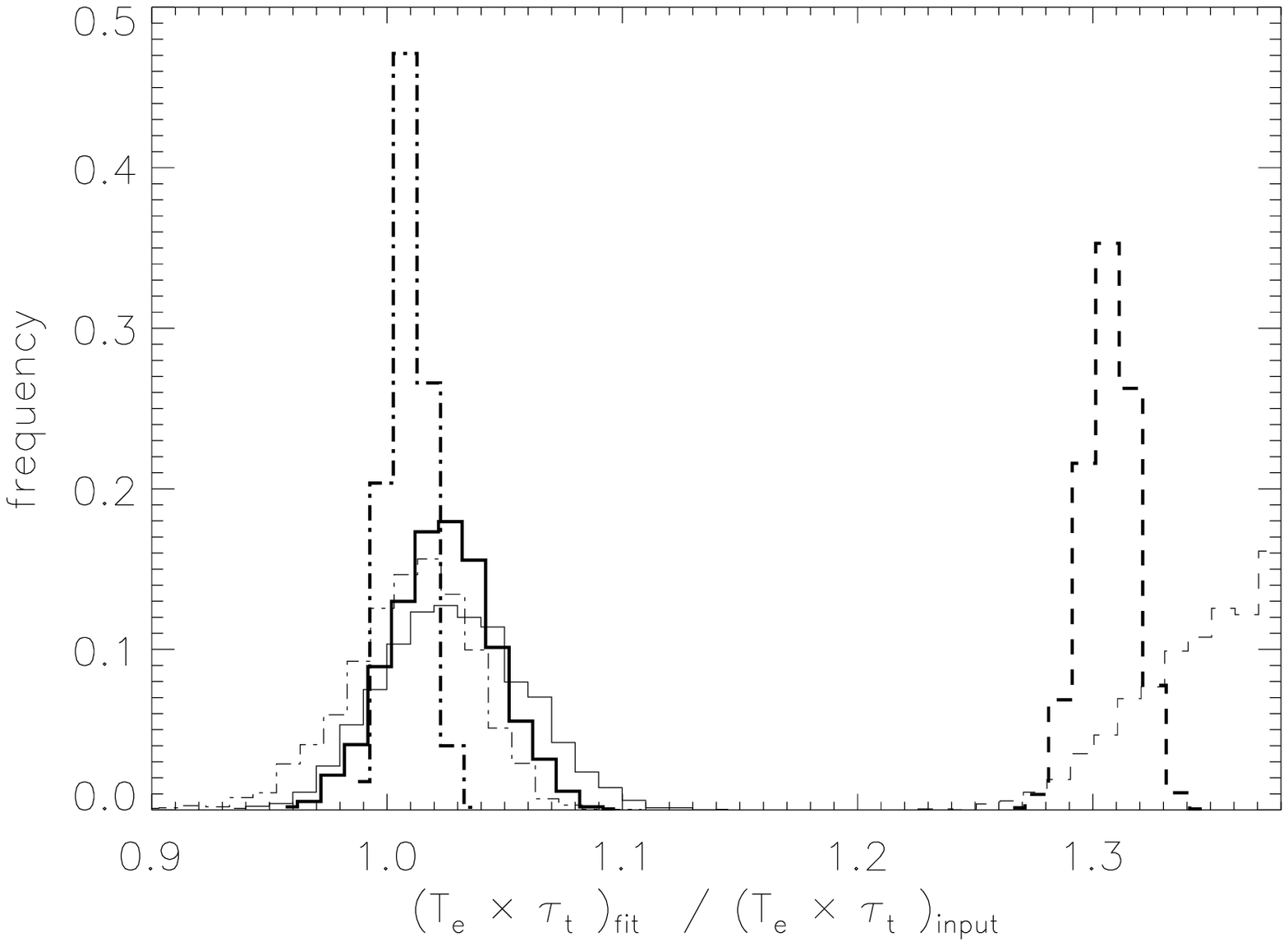}
\caption{ Histograms of the best-fit $T_e \times \tau_t$
normalized to the input of the simulation, for EC0 with three
parameter fits (dashed lines), EC0 with four parameter fits
(continuous lines) and EC1 with four parameter fit (dot-dashed
lines). Thin lines are for a Gaussian prior on $T_e$ centred on
the input value of $T_e$ and with a standard deviation of 8 keV;
thick lines are for a standard deviation of 3 keV.  \label{bias} }
\end{figure}

\begin{figure}
\centering
\includegraphics[width=9cm, angle=90]{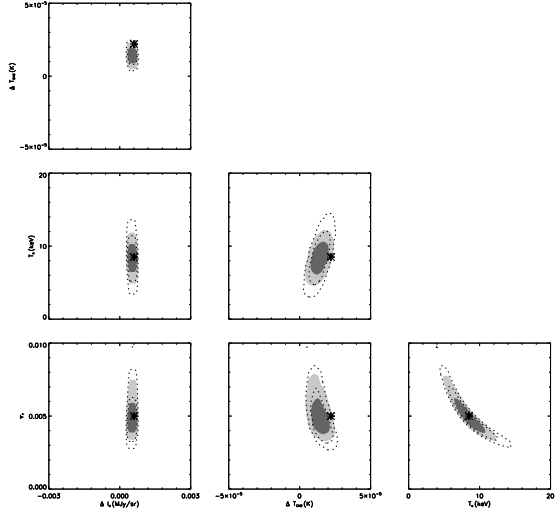}
\caption{ Joint likelihood contours (95.6$\%$ and 68.7$\%$) for
couples of best-fit parameters for the experimental configuration
EC4 (6-band cryogenic photometer operating in L2) where the data
are analysed by fitting four parameters ($\tau_t$, $T_e$, $\Delta
I_d$, $\Delta T_{CMB}$). The filled contours are for a Gaussian
prior on $T_e$ with $\sigma = 3 \ keV$; the dashed contours are
for a Gaussian prior with $\sigma = 8 \ keV$. The * symbols mark
the input values of parameters. \label{fig6} }
\end{figure}

\begin{figure}
\centering
\includegraphics[width=9cm, angle=90]{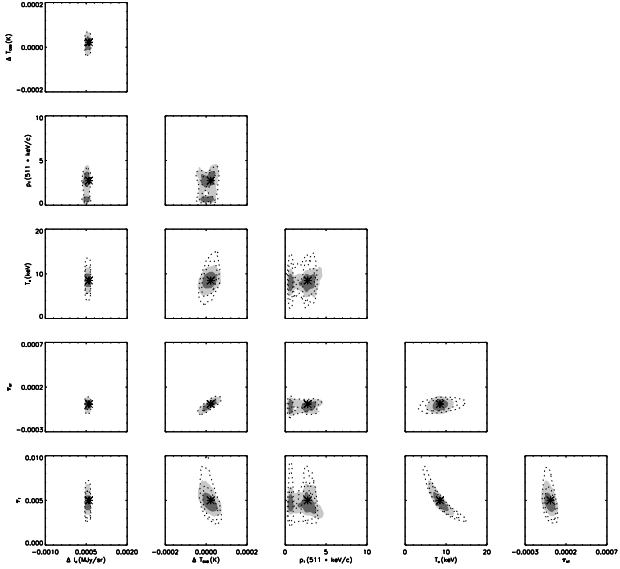}
\caption{ Same as fig. \ref{fig6} (configuration EC4) when the
data are analysed by fitting six parameters ($\tau_t$, $T_e$,
$\Delta I_d$, $\Delta T_{CMB}$, $\tau_{nt}$, $p_1$). \label{fig7}
}
\end{figure}

\begin{figure}
\centering
\includegraphics[width=9cm, angle=90]{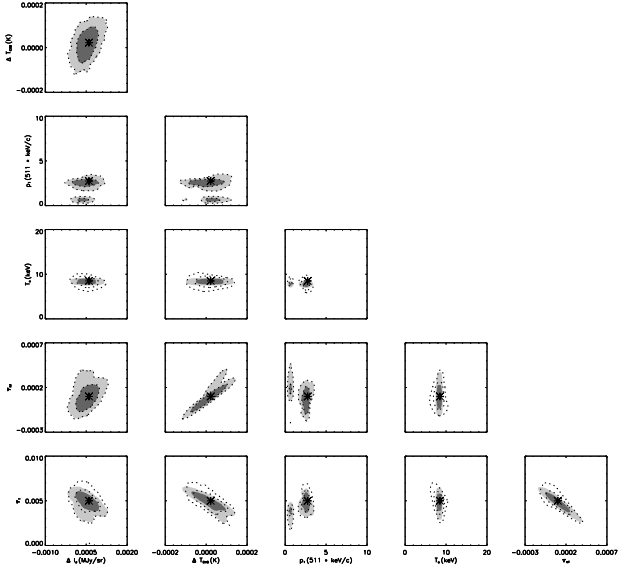}
\caption{ Joint likelihood contours (95.6$\%$ and 68.7$\%$) for
couples of best-fit parameters for the experimental configuration
EC2 (4-range warm spectrometer on a balloon platform).  The filled
contours are for a Gaussian prior on $T_e$ with $\sigma = 3 \
keV$; the dashed contours are for a Gaussian prior with $\sigma =
8 \ keV$. The * symbols mark the input values of parameters.
\label{fig8} }
\end{figure}

In fig. \ref{bias} we plot the histograms of the best-fit $T_e
\times \tau_t$ normalized to the input of the simulation. It is
evident how the bias is reduced passing from EC0 with three
parameters to EC0 with four parameters to EC1, which shows how
important the coverage of high frequencies is, which are difficult
to observe from the ground.

For EC4, where six independent data-sets are available for each
measurement, we tried to fit four parameters ($\tau_t$, $T_e$,
$\Delta I_d$ and $\Delta T_{CMB}$, see fig. \ref{fig6} ) and also
six parameters (including $\tau_{nt}$ and $p_1$, see fig.
\ref{fig7}), always adding the fictitious zero frequency zero
brightness data point. The good coverage of high frequencies and
low photon noise owing to the low radiative background results in
very good performance in terms of statistical errors. Again, the
presence of the non-thermal component produces a bias in the
determination of the other parameters for the 4-parameter fit. In
the 6-parameter fits, the non thermal parameters are basically not
constrained (with a bimodal distribution of the best fit $p_1$),
but the other parameters are well constrained and unbiased. The
effects of parameters degeneracies are still evident, however, and
it is difficult to estimate $\Delta T_{CMB}$ because it tends to
be biased low in the 4-parameter fit, while is not well
constrained in the 6-parameter fits (see also table \ref{tab1}).

The space spectrometers EC2, EC3, and EC5, featuring wide
frequency coverage and low radiative background noise, perform
better, and allow an unbiased recovery of six parameters with
increasing accuracy. If we compare the parameter space volume
constrained by the different configurations, we find that with
respect to EC2, EC3 reduces the volume by a factor 4.6, and EC5 by
a factor 27 (with the 8 keV prior).

The balloon-borne warm spectrometer EC2 is still limited at high
frequencies by radiative background fluctuations. For this reason
$\Delta T_{CMB}$ is basically not constrained (see fig.
\ref{fig8}). However, the other parameters are unbiased and well
constrained even with the very weak prior on $T_e$. An experiment
like OLIMPO, which combines photometric measurements (as in EC1)
{\it and} spectroscopic measurements (as EC2), performing both
during the same flight, can use the first measurement to optimally
constrain $\Delta I_d$ and $\Delta T_{CMB}$, and the second to
better constrain $\tau_t$ and $T_e$. Note that even $\tau_{nt}$ is
close to be detected (and can be detected with an integration time
longer than the 3 hours considered here) and $p_1$ is also
constrained (still with a bimodal distribution). Finally note that
the spectroscopic capabilities allow the user to remove the
contamination of Galactic CO and other cooling lines within the
photometric bands unambiguously.

\begin{figure}
\centering
\includegraphics[width=9cm, angle=90]{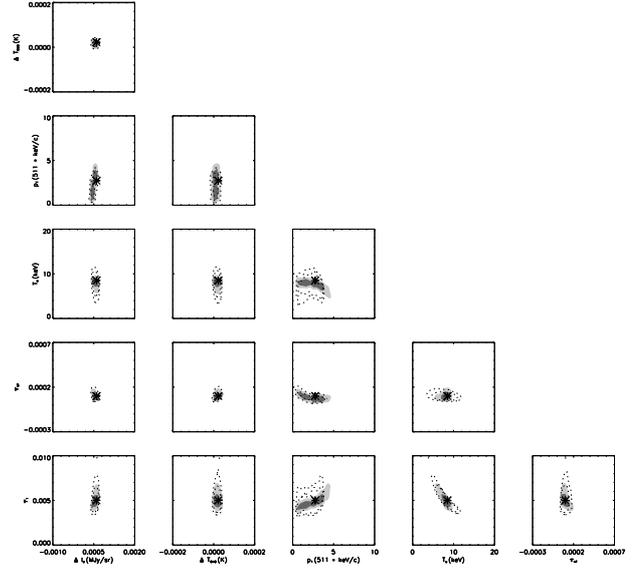}
\caption{ Same as fig. \ref{fig8} for the experimental
configuration EC3 (cold spectrometer on an Earth-orbit satellite
with radiatively cooled (80K) telescope) \label{fig9} }
\end{figure}

The low-Earth-orbit case EC3 improves over the balloon
spectrometer EC2. Cooling the spectrometer results in a
significant unbiased measurement of all six parameters, including
the elusive $\Delta T_{CMB}$ (see fig. \ref{fig9}). Because the
noise is reduced, the effect of degeneracies is more evident. The
3 keV prior is needed to fully exploit the potential of this
configuration.

\begin{figure}
\centering
\includegraphics[width=9cm, angle=90]{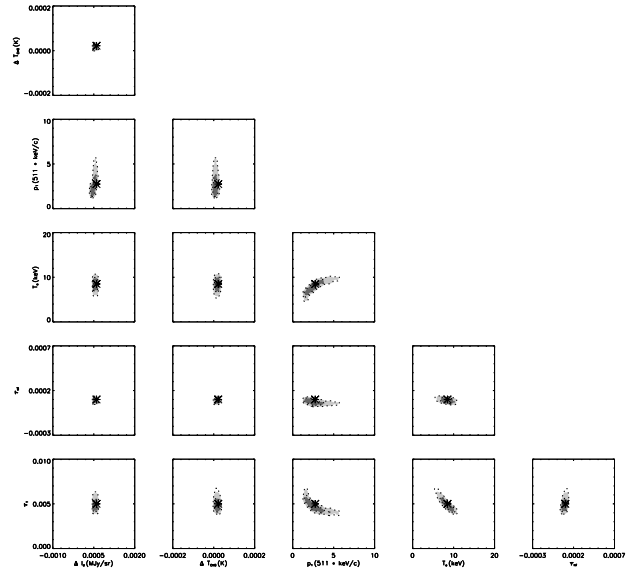}
\caption{ Same as fig. \ref{fig8} for the experimental
configuration EC5 (cold spectrometer on a satellite in L2, with
actively cooled (4K) telescope). \label{fig10} }
\end{figure}

Experimental configuration EC5, where both the spectrometer and
the telescope are cold, results in an improved precision of the
determination of all parameters, and little sensitivity to the
prior on $T_e$ (see fig. \ref{fig10} and table \ref{tab1}).

We have investigated the possibility of extending the frequency
coverage of EC5 up to 1 THz. Keeping the same simplified model for
dust, we found an improvement of about a factor 2 in the
uncertainties of all parameters but $\Delta T_{CMB}$ (which does
not improve) and $\Delta I_d$ (which improves by a factor 5).
However, our simplistic model with a single dust population
probably starts to be insufficient at these high frequencies.
There are, however, other drivers to extend the frequency coverage
of such an ambitious experiment as high as possible: the study of
interstellar atoms and molecules and the study of galaxies in the
redshift desert would benefit significantly from such an
extension.

\section{Conclusions}\label{sec5}

With this simulation-based study we confirmed that few-band
photometric measurements of the SZ effect can be significantly
biased, and that most of the bias is removed by adding bands and
covering high frequencies ($\nu \simgt 240 GHz$). This requires
the use of space-borne experiments. Low-resolution spectroscopic
measurements of the SZ are extremely promising. An exploratory
experiment, featuring a four-band photometer on a balloon and a
plug-in room-temperature FTS [configurations EC1 and EC2, modelled
on the OLIMPO experiment (Masi et al. \cite{Masi08})] is required
to confirm that the expectations estimated here can really be
achieved. In other words, we relied here on the differential
properties of FTS instruments: we need to demonstrate
experimentally that systematic effects are negligible even with
the extreme requirements of SZ measurements. OLIMPO, in the dual
photometric-spectroscopic configuration, is a perfect
demonstrator, and is expected to produce exciting improvements on
the physics of cosmic structures derived from the SZ effect.
Satellite missions can accommodate a cryogenic FTS (a large
cryostat is required anyway to provide the required hold time),
reducing the instrumental background at high frequency. A
Molniya-orbit mission [like the proposed SAGACE (de Bernardis et
al. \cite{debe10})] represents a cost-effective solution, able to
provide an extensive catalogue of unbiased SZ measurements in
various cosmic structures (e.g. galaxy clusters and groups,
radiogalaxy lobes, gaseous halos of galaxies). A cold telescope in
L2 (like the proposed Spectrum-M - millimetron mission) would be
limited only by the intrinsic degeneracy of the parameters, and
would open new horizons in SZ science, allowing a precise
determination of both the thermal and non-thermal components of
the plasma in the cluster.

\begin{table*}[p]
\begin{center}
\begin{tabular}{|c|c|c|c|c|c|c|c|}

\hline Parameter & input & best fit EC0 & best fit EC0 & best fit
EC0 & best fit EC0 & best fit EC1 & best fit EC1
\\
  &  & ground - phot. & ground - phot. & ground - phot. & ground - phot.
& balloon - phot. & balloon - phot.
\\
  &  & prior $\sigma$= 3 keV & prior $\sigma$= 8  keV & prior $\sigma$= 3 keV & prior $\sigma$= 8 keV
     & prior $\sigma$= 3 keV & prior $\sigma$= 8  keV  \\

\hline \hline $\tau_t \times 10^3$ & 5 & $5.2\pm 0.6$  & $3.9\pm
1.3$ & $5.1\pm 0.7$ & $4.7 \pm 3.1$ & $5.0\pm 1.3$ & $5.3 \pm 1.7$
\\

\hline $T(keV)$ & 8.5 & $10.8\pm 1.2$ &  $15.8\pm 3.5$ & $8.7\pm
1.1$  & $12\pm 4$ &  $8.5\pm 0.3$ & $8.6\pm 1.7$
\\

\hline $\Delta T_{CMB} (\mu K) $ & $22$ & - & -  & $19 \pm 6$ & $
11 \pm 6 $ & $16 \pm 2 $ & $16 \pm 4$
\\

\hline $\Delta I_d (Jy/sr)$ & $600$ &  $ -1800 \pm 200 $ &  $
-1100 \pm 500 $ & $390 \pm 480$ & $50 \pm 660$ & $580 \pm 70$ &
$590 \pm 80$
\\
\hline $\tau_{nt} \times 10^3$ & 0.1 & - & - & - & - & - & -
\\
\hline $p_1 (511 \ keV/c)$ & 2.75 & - & - & - & - & - & -
\\
\hline \hline $\langle \ \chi^2 \rangle / DOF $ & - & 6.3/2 &
4.3/2 & 1.0/1 & 1.0/1 & 1.1/1 & 1.1/1
\\

\hline

\end{tabular}
\end{center}

\begin{center}
\begin{tabular}{|c|c|c|c|c|c|}
\hline Parameter & input & best fit EC4 & best fit EC4 & best fit
EC4 & best fit EC4
\\
  &  & L2 - cold phot. & L2 - cold phot. & L2 - cold phot. & L2 - cold phot.
\\
  &  & prior $\sigma$= 3 keV & prior $\sigma$= 8 keV & prior $\sigma$= 3 keV & prior $\sigma$= 8
  keV \\
\hline \hline $\tau_t \times 10^3$ & 5 & $5.2\pm 0.9 $ & $5.4\pm
1.9 $ & $5.0\pm 0.8 $ & $5.3\pm 2.1 $
\\

\hline $T(keV)$ & 8.5  & $8.5\pm 1.5$ & $8.7\pm 2.5$ & $8.6\pm
1.3$ & $9.0\pm 2.7$ \\

\hline $\Delta T_{CMB} (\mu K)$ & $22$ & $16 \pm 4 $ & $16 \pm 4 $
& $21 \pm 22 $ & $22 \pm 23 $
\\

\hline $\Delta I_d (Jy/sr)$ & $600$ & $601 \pm 6 $ & $600 \pm 6$ &
$601 \pm 5$ & $601 \pm 6$ \\

\hline $\tau_{nt} \times 10^3$ & 0.1 & - & - & $0.1 \pm 0.4$ &
$0.1 \pm 0.4$ \\

\hline $p_1 (511 \ keV/c)$ & 2.75  & - & - & $ 6 \pm 4 $ & $6 \pm
4$
\\
\hline \hline $\langle \ \chi^2 \rangle / DOF $ & - & 2.7/3 &
2.4/3 & 1.3/1 & 1.0/1 \\

\hline

\end{tabular}

\end{center}

\begin{center}
\begin{tabular}{|c|c|c|c|c|c|c|c|}
\hline Parameter & input &  best fit EC2 &  best fit EC2 & best
fit EC3 & best fit EC3   & best fit EC5  & best fit EC5
\\
  &   & balloon - warm spec. & balloon - warm spec. & EO - cold spec. & EO - cold spec. & L2 - cold
  spec. & L2 - cold spec.
\\
  &  & prior $\sigma$= 8 keV & prior $\sigma$= 3 keV & prior $\sigma$= 8  keV   & prior $\sigma$= 3 keV & prior $\sigma$= 8
  keV& prior $\sigma$= 3 keV
\\
\hline \hline $\tau_t \times 10^3$ & $5$  & $5.0 \pm 0.9$ & $4.9
\pm 0.8$ & $5.8 \pm 2.6$  & $5.2 \pm 0.6$  & $5.1 \pm 0.6$  & $5.1
\pm 0.5$
\\

\hline $T(keV)$ & 8.5  & $8.4 \pm 0.8$ & $8.5 \pm 0.1$ & $7.7 \pm
2.0$ & $8.1 \pm 0.8$ & $8.5\pm 1.2$ & $8.5\pm 1.0$
\\

\hline $\Delta T_{CMB} (\mu K)$ & $22$ & $20 \pm 50$ & $20 \pm 50$
& $23 \pm 8$ & $22 \pm 8$ & $22 \pm 4$ & $22 \pm 4$
\\

\hline $\Delta I_d (Jy/sr)$ & $600$ & $570 \pm 270$ & $560 \pm
270$ & $590 \pm 40$  & $590 \pm 40$  & $600 \pm 4$ & $600 \pm 4$
\\

\hline $\tau_{nt} \times 10^3$ & 0.1 & $0.1 \pm 0.1$  & $0.1 \pm
0.1$ & $0.12 \pm 0.03$ & $0.11 \pm 0.02$ & $0.10 \pm 0.01$ & $0.10
\pm 0.01$
\\

\hline $p_1 (511 \ keV/c)$ & 2.75 & $2.6 \pm 0.7$ & $2.5 \pm 0.7$
& $2.5 \pm 0.9$ & $2.7 \pm 1.1$ & $3.0 \pm 1.0$ & $2.9 \pm 0.9$
\\
\hline \hline $\langle \ \chi^2 \rangle / DOF $ & -  & 34.9/34 &
34.9/34 & 77.8/78 & 78.0/78 & 110.0/110 & 110.1/110
\\

\hline

\end{tabular}

\end{center}

\begin{center}
\caption{Results of simulations. For each experimental
configuration (EC*) we simulated 3000 measurements and fitted each
simulated measurement using equation \ref{theory}. We report the
averages of the best-fit parameters with their standard deviation.
In the top table we summarize the results for 4-band photometers
for both weak and medium prior for $T_e$; in the middle table we
report the results for a 6-band photometer (Planck-HFI bands). In
the bottom table we report the results for low-resolution ($\Delta
\nu = 6$ GHz) spectrometers with increasing mission complexity
(balloon, Earth orbit, orbit around the Sun-Earth Lagrangian point
L2). \label{tab1}}

\end{center}
\end{table*}

\acknowledgements This work has been supported by Italian Space
Agency contracts ``Millimetron'' and ``OLIMPO'' and by PRIN 2009
``Mm and submm spectroscopy for high-resolution studies of
primeval galaxies and clusters of galaxies'' of the Italian
Ministero dell'Istruzione, dell'Universit\`{a} e della Ricerca.
S.C. acknowledges support by the South African Research Chairs
Initiative of the Department of Science and Technology and
National Research Foundation and by the Square Kilometre Array
(SKA).

\clearpage

\end{document}